\documentclass[12pt]{article}

\usepackage[applemac]{inputenc}
\usepackage[T1]{fontenc}
\usepackage{amsmath}
\usepackage{amsfonts}
\usepackage[english]{babel}

\usepackage{geometry}                
\usepackage[parfill]{parskip}    
\usepackage{graphicx}
\usepackage{amssymb}
\usepackage{epstopdf}
\usepackage[all]{xy}

\usepackage{amsthm} 
\newtheorem{thm}{Theorem}[section] 

\theoremstyle{definition} 
 
\theoremstyle{axiom}

\theoremstyle{remark} 
 
\theoremstyle{plain}

\theoremstyle{plain}

\begin{document}

\title{Some remarks on history and pre-history of Feynman path integral}
\author{Daniel Parrochia}
\date{University of Lyon (France)}
\maketitle

\textbf{Abstract.}
One usually refers the Feynman's concept of path integral to the work of Norbert Wiener on Brownian motion in the early 1920s. This view is not false and we show in this article that Wiener used the first path integral of the history of physics to describe the Brownian motion. That said, Wiener, as he pointed out, was inspired by the work of some French mathematicians, particularly Gâteaux and Levy. Moreover, although Richard Feynman has independently found this notion, we show that in the course of the 1930s, while searching a kind of geometrization of quantum mechanics, another French mathematician, Adolphe Buhl, noticed by the philosopher Gaston Bachelard, had himself been close to forge such a notion. This reminder does not detract from Feynman's remarkable discovery, which must undeniably be attributed to him. We also show, however, that the difficulties of this notion had to wait many years before being resolved, and it was only recently that the «path integral» could be rigorously established from a mathematical point of view.


\section{Feynman path integral}

In quantum mechanics, in the Schrödinger representation, the evolution in time of a quantum system is characterized (at the infinitesimal level) by the Hamiltonian operator, as expressed by the famous Schrödinger equation:
\[
i\hbar {\frac  {d|\psi \rangle }{dt}}={\hat  {H}}|\psi \rangle 
\]
where $|\psi \rangle$ is the wave function of the system, and $\hat {H}$ is the Hamiltonian operator. In a stationary state, we know that :
\[
|\psi (t)\rangle =e^{{-i{\frac  {Et}{\hbar }}}}|\psi (0)\rangle ,
\]
where $E$ is the energy of this stationary state. 

The form of this equation, which implies, as solutions, multiple wave functions and their combinations, leading  to the problematic concept of «superposition of states», could have been a motivation, for a young scientist, to search another formulation of non relativistic quantum theory. But it is not the case. 

In the 1940s, when Richard Feynman (1918-1988) sought to reformulate quantum mechanics from a very general variational principle - the principle of least action - his original motivation was quite different. If he chose to start not from the Hamiltonian operator of the wave equation but from the Lagrangian, it stemmed from the will to obtain a quantum-mechanical formulation for the so-called "Wheeler-Feynman absorber theory", a device which tried to justify an approach of the electrodynamics without the notion of field. As T. Sauer has shown, such an idea was wrong, and this particular context has all interest to be forgotten (see \cite{Sau}, 8-9). However, reformulating quantum mechanics by starting from the Lagrangian instead of the Hamiltonian, was a promising idea. 

\subsection{First attempt}

To tell the truth,  Feynman was not the first to embark in such an attempt. This approach was first developed by P. A. M. Dirac in the years 1932-33. According to Dirac, quantum mechanics was built up on a foundation of analogy with the Hamiltonian theory of classical mechanics. But there was some reasons to believe that the Lagrangian was more fundamental : first, there was no action principle in the Hamiltonian theory; second, the Lagrangian method could easily be expressed relativistically, while the Hamiltonian one was essentially non relativistic (see \cite{Dir}, 63).

In 1933, Dirac took a few steps in the direction of such a reformulation. In a 9-page article published in an obscure Soviet scientific journal, he laid the foundation for what would one day become Feynman's theory of path integrals.

Comparing classical mechanics and quantum mechanics, Dirac observed, in particular, that, for two neighboring instants $t$ and $t + \epsilon$, the elementary transition amplitude $\langle q_ {2} (t + \epsilon) | q_ {1} (t) \rangle$ was analogous to $\exp (iS [q] / \hbar)$ (see \cite{Dir}, 69).

In this formula, the magnitude $S[q(t)]$ is the classical action:
\[
S[q_{2}(t+\epsilon ),q_{1}(t)]\ =\ \int _{t}^{t+\epsilon }L(q,{\dot {q}})\ \mathrm{d} t\ =\ L\left(q_{1},{\frac {q_{2}-q_{1}}{\epsilon }}\right)\epsilon.
\]

Seeking information on the attempts already made to build up quantum mechanics from the lagrangian, Feynman had been led to read, apparently on the advice of his friend Jehle (see \cite{Sch}), the 1933 article by Dirac, and, for sure, did not fail to notice Dirac's sentence and the analogy he pointed out.

In order to understand what Dirac means by «analogous», he studied the case of a non-relativistic particle of mass $m$, for which the Lagrangian is :

\[
L (q, {\dot {q}}) \ = \ {\frac {m} {2}} {\dot {q}} ^ {2} \ - \ V (q)
\]
Knowing that :
\[
 \langle q_ {2} | \psi (t + \epsilon) \rangle \ = \ \psi (q_ {2}, t + \epsilon) \ = \ \int \mathrm{d} q_ {1} \, \langle q_ {2} (t + \epsilon) | q_ {1} (t) \rangle \, \langle q_ {1} | \psi (t) \rangle
\]

Feynman then assumed a relationship of proportionality:

\[
 \psi (q_ {2}, t + \epsilon) \ = \ A \ \int \mathrm {d} q_ {1} \, \exp \, \left (\, i \, {\frac {S [q (t)]} {\hbar}} \, \right) \ \psi (q_ {1}, t)
\]

where $A$ is an unknown constant. In the presence of Jehle, Feynman demonstrated that this equation implies that $\psi (q, t)$ obeys the Schrödinger equation:

\[
\left[\,-\ {\frac {\hbar ^{2}}{2m}}\ {\frac {\partial ^{2}~~}{\partial q^{2}}}\ +\ V(q)\,\right]\ \psi (q,t)\ =\ i\,\hbar \ {\frac {\partial ~~}{\partial t}}\psi (q,t)
\]

provided that the unknown constant $A$ is equal to:

\[
A\ =\ {\sqrt {\frac {m}{2\pi \hbar it}}}.
\]

All this was happening in the 1940s, that is, before Feynman supported his 1942 thesis under the direction of John Archibald Wheeler, and far before the first public formulation of path integral theory in 1948 (see \cite{Fey1}).

\subsection{Main presentation}

The general case has been exposed in the article of 1948, but we prefer to report here the clear formulation of 1965. In order to introduce very pedagogically the concept of path integral, Feynman and Higgs, in their book, started from a Young double slit type experiment where a source sends electrons on a plate pierced with two slots before they then reach a detection screen. They suppose then one introduces a couple of extra-screens $E$ and $D$ between the source and the holes, so that, in each of them, one drills a few holes respectively called $E_{1}... E_{2}$ and $D_{1}... D_{2}$, as in Fig. 1. For simplicity, they assume the electrons are constrained to move in the $xy$ plane. Then, as they said, «there are several alternative paths which an electron may take in going from the source to either hole in screen $C$. It could go from the source to $E_{2}$, and then $D_{3}$, and then the hole 1; or it could go from the source to $E_{3}$, then $D_{1}$, and finally the hole 1; etc. Each of these paths has its own amplitude. The complete amplitude is the sum of all of them.»(\cite{Fey2}, 20).

\begin{figure}[h] 
	   \centering
	      \vspace{-2\baselineskip}
	   \includegraphics[width=4in]{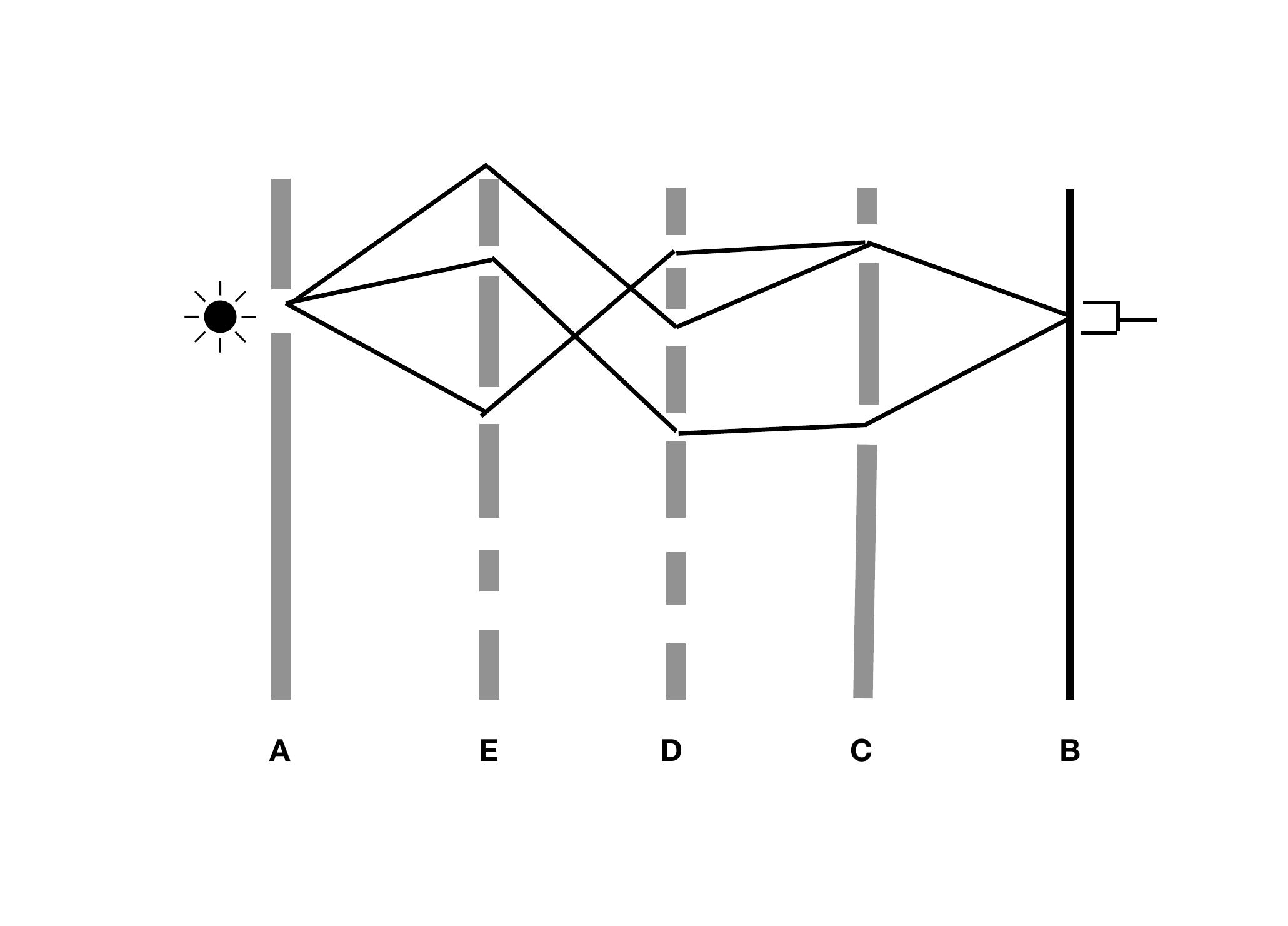} 
	      \vspace{-2\baselineskip}
	   \caption{Young's experiment with 2 interpolated screens.}
	   \label{fig: Feyn11}
	\end{figure}
	
	Then Feynman and Higgs continue the process : «suppose we continue to drill holes in the screens $E$ and $D$ until there is nothing left of the screens. The path of an electron must be now specified by the height $x_{E}$ at which the electron passes the position $y_{E}$ at the nonexistent screen $E$, together with the height $x_{D}$ at the position $y_{D}$ (see Fig. 2). To each pair of heights correspond an amplitude. The principle of superposition still applies, and we must take the sum (or by now, the integral) of these amplitudes over all possible values of $x_{D}$ and $x_{E}$.
	
	\begin{figure}[h] 
	   \centering
	      \vspace{-2\baselineskip}
	   \includegraphics[width=4in]{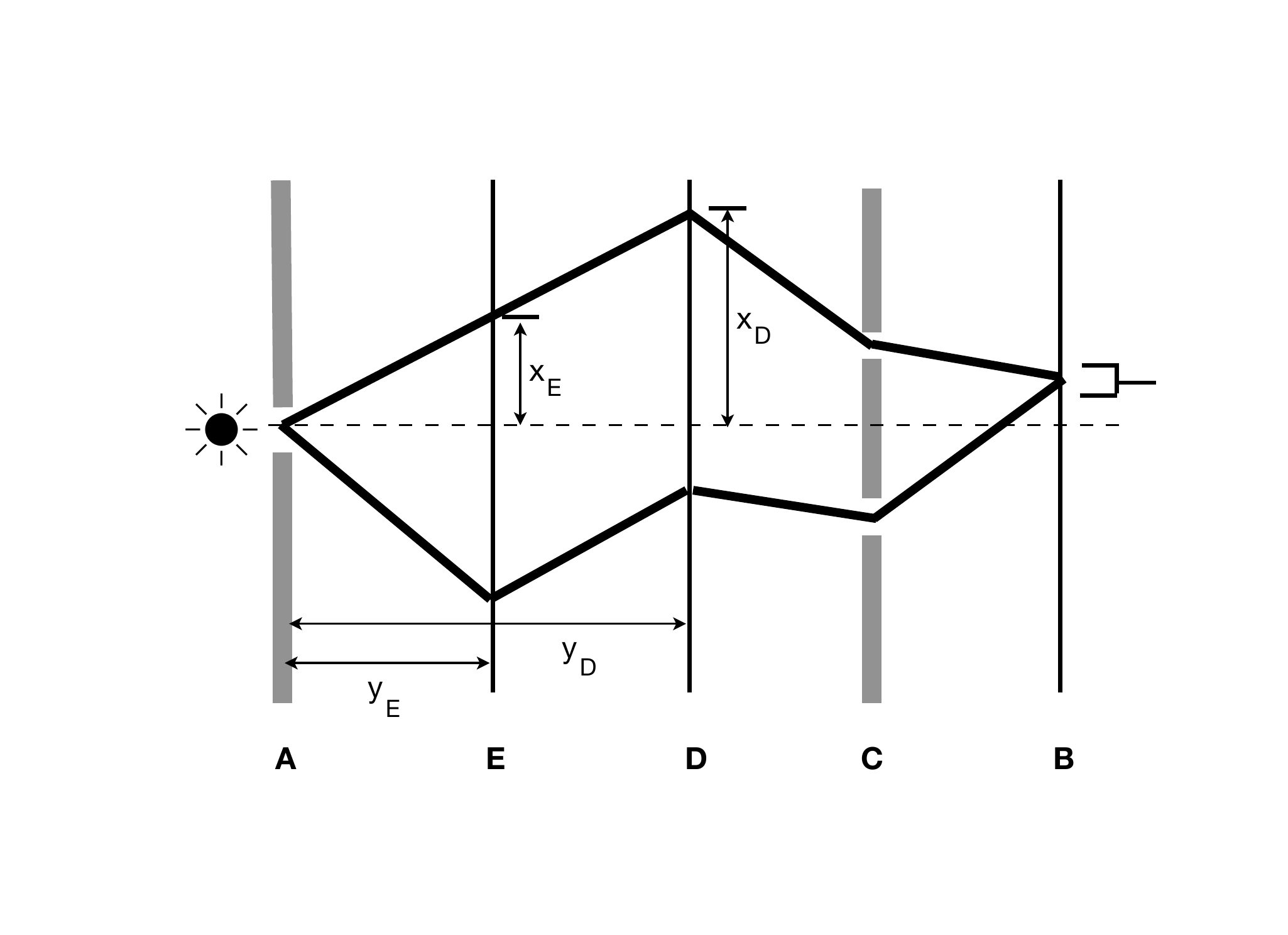} 
	      \vspace{-2\baselineskip}
	   \caption{Young's experiment with many interpolated screens.}
	   \label{fig: Feyn21}
	\end{figure}

	Clearly, the next thing to do is to place more and more screens between the source and the hole 1, and in each screen drill so many holes than there is nothing left. Throughout this process we continue to refine the definition of the path of the electron, until we arrive at the sensible idea that a path is merly height as a particular function of distance, or $x(y)$. We also continue to apply the principle of superposition, until we arrive at the integral over all paths of the amplitude for each path.»(\cite{Feyn2}, 

More physically, it is introduced very naturally from considerations on classical mechanics. In classical mechanics, the position of a particle at any time can be specified by a coordinate $x$, which is a function of time.  By a «path», Feynman and Higgs mean just the function $x(t)$. Now if a particle at an initial time $t_{a}$ starts from point $x_{a}$ to reach a final point $x_{b}$ at time $t_{b}$, one simply says that the particle goes from $a$ to $b$ and the function $x(t)$ will have the property that  $x_{t_{a}} = x_{a}$ and $x_{t_{b}} = x_{b}$. In quantum mechanics, one will have an amplitude (or kernel) $K(a, b)$, to get from the point $a$ to the point $b$, and this will be the sum over all trajectories between the end points $a$ and $b$ of a contribution for each. This is the great contrast with classical mechanics where there is only one specific and particular trajectory which goes from $a$ to $b$. How to prove that?

The authors start from the {\it principle of least action}, which is the condition that determines a particular path $\bar{x}(t)$ out of all the other possible paths. This path $\bar{x}(t)$ is the one for which a certain quantity $S$ is a minimum ($S$ is, in fact, an {\it extremum}, i.e. its value remains unchanged in the first order if the path $\bar{x}(t)$ is slightly modified).

The quantity $S$ is given by the expression:
\begin{equation}
S = \int_{t_{a}}^{t_{b}} L(\dot{x}, x, t)dt,
\end{equation}
where $L$ is the Lagrangian of the system. For a particle of mass $m$ subject to a potential energy $V(x, t)$, which is a function of position and time, the lagrangian is :
\begin{equation}
L = \frac{m}{2}\dot{x}^2 - V(x, t).
\end{equation}
As we know, the form of the extremal paths $\bar{x}(t)$ is determined through the procedures of the calculus of variations, e.g. if we suppose the path is varied away from  $\bar{x}(t)$ by an amount $\delta x(t)$, the conditions that the end points are fixed require :
\[
\delta x(t_{a}) = \delta x(t_{b} = 0,
\]
and the condition that  $\bar{x}(t)$ be an extremum of $S$ means :
\[
\delta S = S[\bar{x} + \delta x] - S[\bar{x}] = 0,
\]
to first order in $\delta x$. So, using the definition (9), we may write :
\[
 S[\bar{x} + \delta x] = \int_{t_{a}}^{t_{b}} L(\dot{x} + \delta \dot{x}, x + \delta x, t)dt,
\]
\[
= \int_{t_{a}}^{t_{b}} \left[L(\dot{x}, x, t) + \delta \dot{x} \frac{\partial L} {\partial \dot{x}} + \delta x \frac{\partial L}{\partial x}\right] dt,
\]
\begin{equation}
= S[x] + \int_{t_{a}}^{t_{b}} \left[\delta \dot{x} \frac{\partial L} {\partial \dot{x}} + \delta x \frac{\partial L}{\partial x}\right] dt.
\end{equation}
Upon integration by parts, the variation in $S$ becomes :
\[
\delta S = \left[\delta x \frac{\partial L}{\partial \dot{x}}\right]_{t(a)}^{t_{b}} - \int_{t_{a}}^{t_{b}} \delta x  \left[ \frac{d}{dt}\left(\frac{\partial L} {\partial \dot{x}}\right) - \frac{\partial L}{\partial x}\right] dt.
\]

Since $\delta(x)$ is zero at the end points, the first term of the right hand-side of the equation is zero. But between the end points, $\delta x(t)$ can take on any arbitrary value. Thus the extremum is that curve along which the following condition is always satisfied:
\[
\frac{d}{dt}\left(\frac{\partial L}{\partial \dot{x}}\right)-\frac{\partial L}{\partial x} = 0, 
\]
this being the classical lagrangian equation of motion.

In classical mechanics, the whole form of the action integral $S = \int Ldt$ is interesting, not only its extremum value, because we want to know the action along a set of neighboring paths, in order to determine the path of least action.

In quantum mechanics, both the form of the integral and the value of the extremum are important. In this new context, we must say how much each trajectory contributes to the total amplitude to go from $a$ to $b$. It is not just the particular path of extreme action which contributes. In fact, all paths contribute. They contribute equal magnitude to the total amplitude, but contribute at different phases, each of them being the action of $S$ for a given path, expressed in units of the quantum of action $\hslash$.

The probability $P(b, a)$ to go from a point $x_{a}$ at time $t_{a}$ to a point $x_{b}$ at time $t_{b}$ is the square $P(a, a) = |K(b, a]^2$ or an amplitude $K(b, a)$ to go from $a$ to $b$. This amplitude is the sum of contributions $\phi[x(t)]$ from each path. We have :
\[
K(b, a) = \sum_{\mathrm{path\ from\ a\ to\ b}} \phi[x(t)].
\] 
As Feynman has proved (see above), the contribution of a path has a phase proportional to the action $S$ :
\[
\phi[x(t)] = e^{(i/ \hslash)S[x(t)]}.
\]
The action is that for the corresponding classical system and the constant will be chosen to normalize $K$ correctly.

On the scale where classical physics might be expected to work, $S >> \hslash$ and the trajectories converge to the one define by the principle of less action. But in quantum mechanics, i.e at an atomic level, $S$ may be comparable to $\hslash$, so all trajectories must be added in details (see Fig. 3) and none of them is of overwhelming importance. 

\begin{figure}[h] 
	   \centering
	      \vspace{-2\baselineskip}
	   \includegraphics[width=4in]{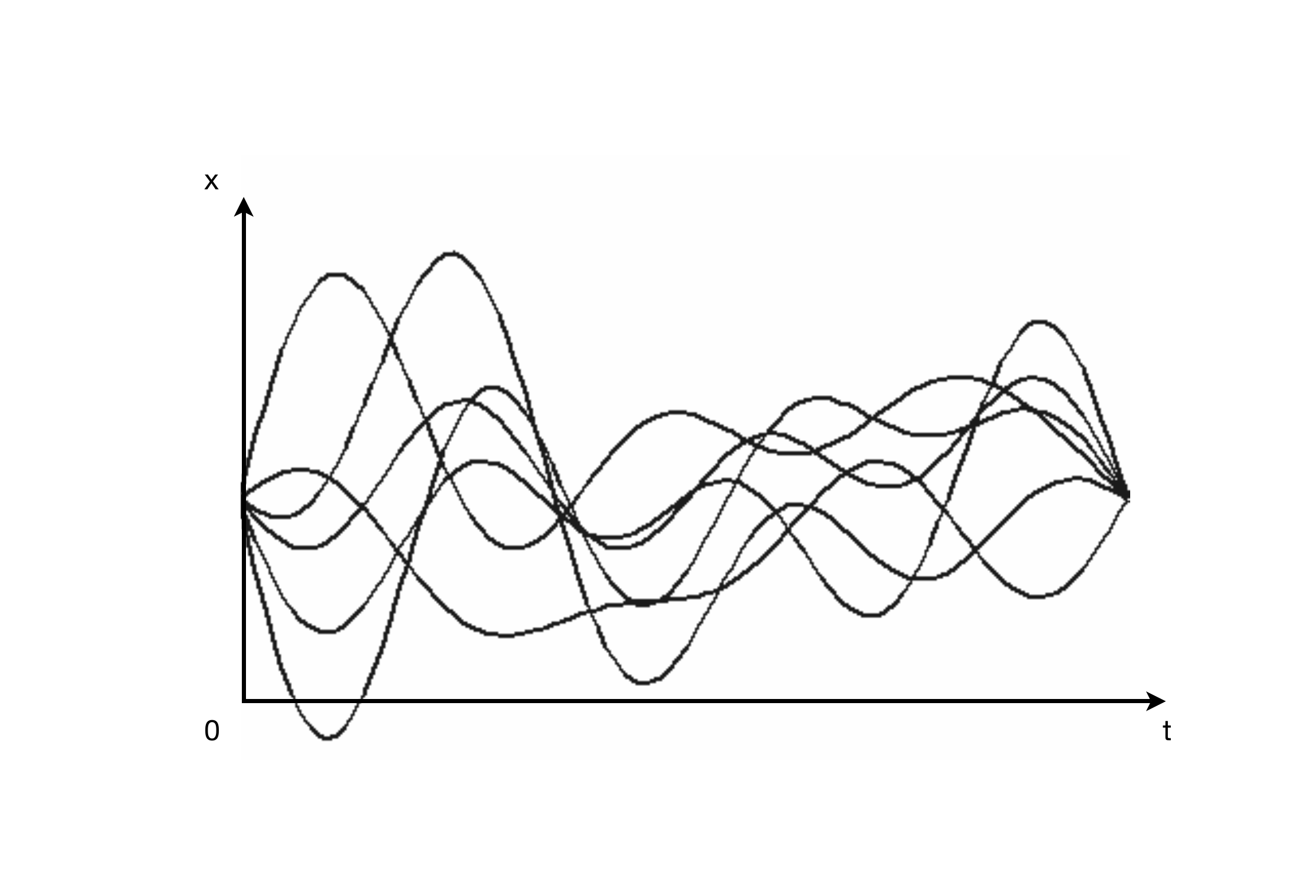} 
	      \vspace{-2\baselineskip}
	   \caption{A subset of the infinite set of trajectories}
	   \label{fig: Feynman1}
	\end{figure}
	
The problem is that the number of paths is of higher order of infinity and it is not obvious what measure is to be given to the space of paths.

In analogy with the Riemann integral where the area $A$ under the curve is the sum of  all its ordinates, such the sum $\sum f(x_{i})$ appears to be proportional to $A,\ A$ being the limit of the sum when the interval $h$ between two successive ordinates tends to zero, Feynman and Higgs construct their sum by chosing first a finite subset of all paths. To do this, they divide the independent variable time into steps of width $\epsilon$, which gives a set of values $t_{i}$ spaced an interval $\epsilon$ apart between the values $t_{a}$ and $t_{b}$, so that, at each time, is selected some special point $x_{i}$. They then construct a path by connecting all the points so selected with straight lines. Thus it is possible to define a sum over all paths connected in this manner by taking a multiple integral over all values of $x_{i}$, for $i \in [1, N-1]$, where:
\[
N\epsilon = t_{b} - t_{a},
\]
\[
\epsilon = t_{i+1} - t_{i}
\]
\[
t_{0} = t_{a} \qquad \qquad t_{N} = t_{b},
\]
\[
x_{0} = x_{a} \qquad \qquad x_{N} = x_{b},
\]
the resulting equation being:

\begin{equation}
K(b, a) \sim \int ...  \int \int \phi[x(t)] d_{x_{1}} d_{x_{2}} \ ... \ d_{x_{N-1}} .
\end{equation}

One does not integrate over $x_{0}$ and $x_{N}$ because they are the fixed points $x_{a}$ and $x_{b}$. As in the case of the Riemann sum, the convergence is expected by making $\epsilon$ smaller and by finding a normalizing factor which depends on it. Unfortunately, as Feynman and Higgs recognize, one does know how to do that in general. But  «we do know how to give the definition for all situations which so far seem to have a practical value» (\cite{Fey2}, 33). For example, in the case the lagrangian is given by (10), the normalizing factor is $A^{-N}$ where:
\[
A = \left(\frac{2\pi \hslash\epsilon}{m}\right)^{1/2}.
\] 
With this factor, the limit exist and we can write:
\[
K(b, a) = \lim_{\epsilon \to 0} \frac{1}{A} \int  ... \int \int e^{(i/\hslash) S[b, a]}\ \frac{dx_{1}}{A}\ \frac{dx_{2}}{A}\ ... \ \frac{dx_{N-1}}{A}.
\]  

The least we can say is that this way of constructing the path integral is not mathematically correct, and it will be a long time before it becomes so.

\section{Wiener on Brownian motion}

Before going further, let us take a glance at the history of path integral. Feynman's approach, in fact, was not the first of its kind. One used to say that the basic idea of the path integral formulation can be traced back to Norbert Wiener, who familiarized the Wiener integral for solving problems in diffusion and Brownian motion. 

It is a fact that the apparently irregular motion known as «Brownian motion», however non deterministic it may be, still obeys certain rules. The foundations of a strict theory of Brownian motion were developed in the pioneering work by A. Einstein as soon as 1905-1906 (see \cite{Ein1}, \cite{Ein2}). This lead to diffusion equations and their solutions, and also to the definition of some properties of transition probability. 

\subsection{Einstein's diffusion equation}

Let us start with a large number of particles which perform Brownian motion along some axis (for simplicity, we shall only consider one-dimensional movement), and which do not interact with each other. Let $w(x, t)$ the number of particles in a small interval $dx$ around the position $x$ at a time $t$ (the density of particles) and let $j(x, t)$ denote the particle current, i.e. the net number of Brownian particles that pass the point $x$ in the direction of increasing value of $x$ per unity of time. It is an experimental fact that the particle current is proportional to the gradient of their density. We have :
\begin{equation}
j(x, t) = - D \frac{\partial w(x, t)}{\partial x},
\end{equation}
where $D$ is the diffusion constant. 

If particles are neither created nor destroyed, the density and the current obey the continuity equation,
\begin{equation}
\frac{\partial w(x, t)}{\partial t} = - \frac{\partial j(x, t)}{\partial x}.
\end{equation}
which, due to (5), can also be written in the form:
\begin{equation}
\frac{\partial w(x, t)}{\partial t} = - D  \frac{\partial^2 w(x, t)}{\partial^2 x}.
\end{equation}
This is the well-known equation diffusion whose an analogous derivation may be carried out for a particle wandering in a space of arbitrary dimension $d$.
The solution of (7) may be obtained by the means of Fourier transform and its representation of the Dirac $\delta$-function, giving finally:
\begin{equation}
w(x, t) = \frac{1}{\sqrt{4\pi Dt}} \ \mathrm{exp}\left\{-\frac{x^2}{4Dt}\right\}.
\end{equation}

By construction, this distribution is the solution of the diffusion equation (3), when the initial condition is that $w(x, t) \rightarrow \delta(x)$ when $t \rightarrow 0$ ($\delta(x)$ being the Dirac $\delta$-function).

\subsection{Wiener's path integral}

In the 1920s, Wiener (see \cite{Wie1},  \cite{Wie2}, \cite{Wie3}, \cite{Wie4}) introduced a new concept of integral that will precisely become, several years after, the notion of «path integral». At this time, as he noted in his third paper on the subject, integration in infinitely many dimensions was a relatively little-studied problem : «Apart from certain tentative investigations of Fréchet and E. H. Moore, he wrote, practically all that has been done on it is due to Gâteaux, Levy, Daniell and the author»(\cite{Wie3}, 132). Gâteaux and Levy have particularly inspired him and the later received in this article full credit. 

To go a little faster, we shall not follow here in details the development of Wiener and shall prefer recall his main results through the presentation of the XX$^th$ century's well-known textbook from Chaichian and Demychev.

Consider, for simplicity, a one dimensional Brownian motion ending in a certain interval $AB$ (see Fig. 4). The probability of a brownian particle to be, at the moment $t$, anywhere in $AB$ is given by:
\begin{equation}
\mathbb{P} \{x(t) \in [AB] \} = \int_{A}^{B} dx\ w(x;t).
\end{equation}
where $w(x,t)$ is the probability density to find the particle in $x$ at $t$.

\begin{figure}[h] 
	   \centering
	      \vspace{-4\baselineskip}
	   \includegraphics[width=4in]{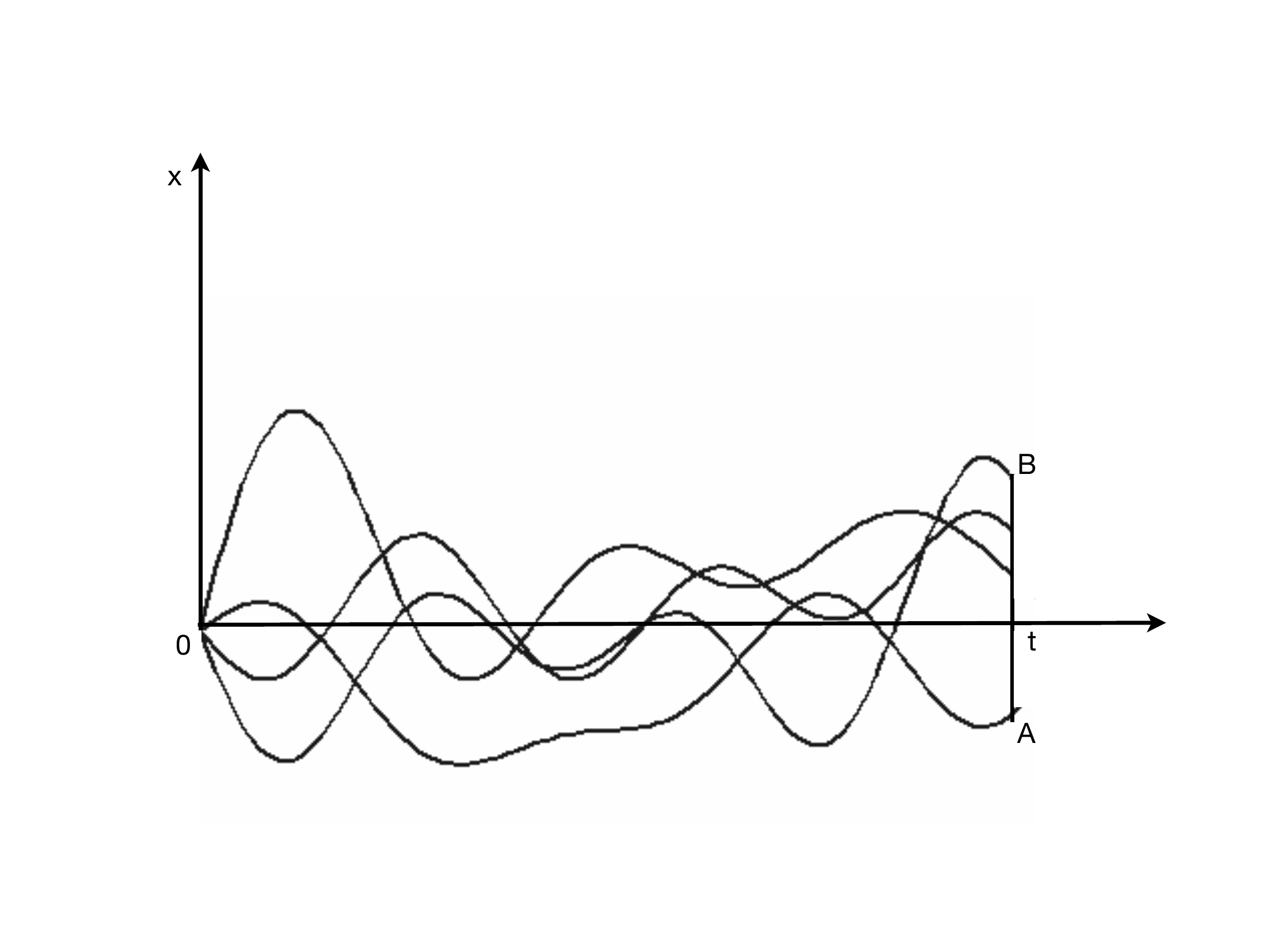} 
	      \vspace{-2\baselineskip}
	   \caption{Multiple trajectories of a Brownian particle starting at the origin and ending somewhere in $AB$.}
	   \label{fig: Wiener11}
	\end{figure}

As Chaichian and Demichev have shown (\cite{Cha}, 22), the stochastic process contains more information than just knowing the set of probabilities $\mathbb{P} \{x(t) \in [AB] \}$. In fact, the main characteristic of such a process is to be a probability of a compound event. During the Brownian motion, the particle, starting at $x(0) = 0$, successively passes through the gates $A_{1} \leq x(t_{1}) \leq B_{1}, A_{2} \leq x(t_{2}) \leq B_{2}, ... , A_{N} \leq x(t_{N}) \leq B_{N}$ at the corresponding time $t_{1}, t_{2}, ... , t_{N}$, as shown in Fig. 5.

\begin{figure}[h] 
	   \centering
	      \vspace{-2\baselineskip}
	   \includegraphics[width=4in]{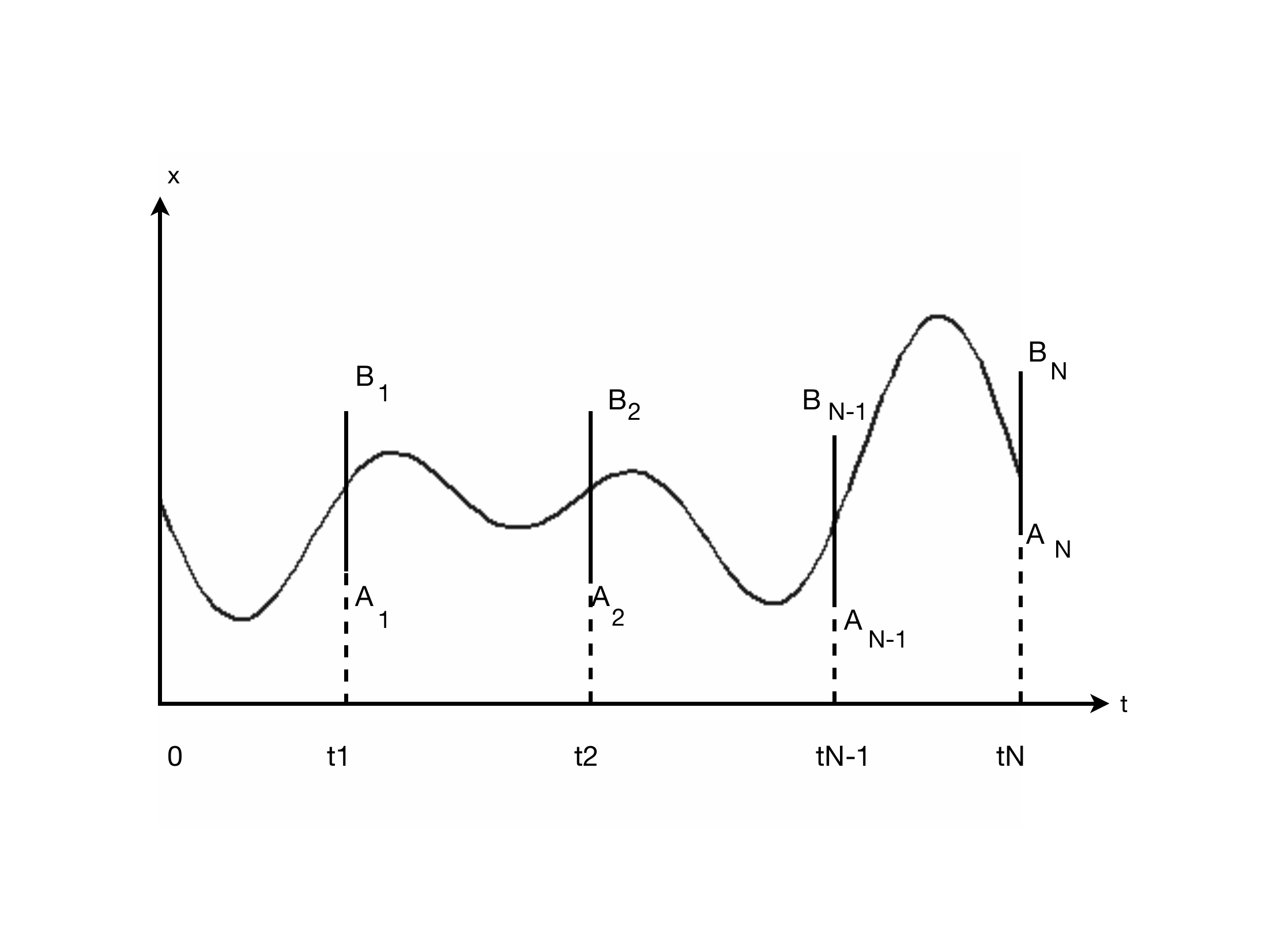} 
	      \vspace{-2\baselineskip}
	   \caption{A trajectory of a Brownian particle starting $t = 0$ and passing through the gate $A_{i}B_{i}$ at the times $t_{i}\ (i = 1,... N)$.}
	   \label{fig: Wiener21}
	\end{figure}

As the subsequent displacements of the Brownian particle are statistically independent (Markovian property), we have necessarily multiplicative probabilities, and so:
\[
\mathbb{P} \{x(t_{1}) \in [A_{1}B_{1}], x(t_{2}) \in [A_{2}B_{2}], ... , x(t_{N}) \in [A_{N}B_{N}] \}
\]
\[
= \mathbb{P} \{x(t_{1}) \in [A_{1}B_{1}]\} \times \mathbb{P} \{x(t_{2}) \in [A_{2}B_{2}]\} \times ... \times \mathbb{P} \{x(t_{N}) \in [A_{N}B_{N}]\}
\]
\begin{equation}
 = \int_{A_{1}}^{B_{1}} dx_{1}\frac{\mathrm{exp}\left\{-\frac{x^2_{1}}{4Dt}\right\}}{\sqrt{4\pi Dt_{1}}} \times  \int_{A_{2}}^{B_{2}} dx_{2}\frac{\mathrm{exp}\left\{-\frac{(x_{2}-x_{1})^2}{4D(t_{2}-t_{1})}\right\}}{\sqrt{4\pi D(t_{2}-t_{1})}}
\end{equation}
\[
\times  \int_{A_{3}}^{B_{3}} dx_{3}\frac{\mathrm{exp}\left\{-\frac{(x_{3}-x_{2})^2}{4D(t_{3}-t_{2})}\right\}}{\sqrt{4\pi D(t_{3}-t_{2})}}... \times  \int_{A_{N}}^{B_{N}} dx_{N}\frac{\mathrm{exp}\left\{-\frac{(x_{N}-x_{N-1})^2}{4D(t_{N}-t_{N-1})}\right\}}{\sqrt{4\pi D(t_{N}-t_{N-1})}}.
\]

Now to reach the limit of the continuous time, one has to diminish the sizes of each gate and to infinitely increase their number, so that:

\[
t_{i}-t_{i-1} \equiv \Delta t_{i} \rightarrow 0  \qquad \qquad 1 \leq i \leq N.
\]

The position $x(t)$ of a particle depends on the continuous time variable and we get a {\it stochastic process} with independent increment and no memory, i.e. a {\it Markov process}. In general, there is no restriction in such processes either on the initial distribution or in the transition probabilities. But in the case of Brownian motion, there are particular conditions which lead to (10), so the process is said to be a {\it Wiener process}.

We can now consider the continuous limit in (10), in order to get the probability that the Brownian particle moves through an infinite number of infinitesimal gates $dx$ along the trajectory $x(t)$. And so we have:
\[
\lim_{\substack{\Delta t_{i} \rightarrow 0  \\ 
N \rightarrow \infty} }
\mathrm{exp} \left\{- \sum_{i=1}^N \frac{(x_{i} - x_{i-1})^2}{4D(t_{i}-t_{i-1})}\right\} \prod_{i = 1}^N \frac{dx_{i}}{\sqrt{4\pi D(t_{i}-t_{i-1})}}
\]
\begin{equation}
= \lim_{\substack{\Delta t_{i} \rightarrow 0  \\ 
N \rightarrow \infty} }
\mathrm{exp} \left\{- \frac{1}{4D} \sum_{i=1}^N \left(\frac{x_{i} - x_{i-1}}{t_{i}-t_{i-1}}\right)^2 \Delta t_{i}\right\} \prod_{i = 1}^N \frac{dx_{i}}{\sqrt{4\pi D \Delta t_{i})}}
\end{equation}
\[
\equiv \mathrm{exp} \left\{- \frac{1}{4D} \int_{0}^t d\tau \dot{x}^2(\tau)\right\} \prod_{\tau = 0}^t \frac{dx(\tau)}{\sqrt{4\pi D d\tau}}.
\]

 In other words, one obtains the probability of the particle motion inside the infinitesimally thin tube surrounding the path $x(\tau)$ or, more simply, moving along the trajectory $x(\tau)$.

Let us denote $\mathcal{C}[x_{1}, t_{1}, \mathcal{B}, t_{2}]$ the set of trajectories, starting at the point $x_{1} = x(t_{1})$ at the time $t_{1}$, and having the endpoint $x_{2} = x(t_{2})$ at the time $t_{2}$, in some domain $\mathcal{B}$ of $\mathbb{R}^d$.

For example, $\mathcal{C}[x_{1}, t_{1}, AB, t_{2}]$ will denote, in the one-dimensional case, the set of trajectories, starting at the point $x_{1} = x(t_{1})$ at the time $t_{1}$, and ending at the gate $[AB]$ at the time $t_{2}$.

 If a trajectory has an arbitrary endpoint in the interval ]$-\infty, +\infty$[ for all coordinates, then we shall simplify this notation by omitting the indication of the whole space $\mathbb{R}^d$ and just write : $\mathcal{C}[x_{1}, t_{1}, t_{2}]$.

These notations are applicable to spaces or arbitrary dimensions but we will continue now, for simplicity, to consider the one-dimensional case which contains all the essential points for a path integral description of Brownian motion in spaces of higher dimensions.

It is clear that to obtain the probability that the particle ends up somewhere in the gate $[AB]$ at the time $t$, we have to sum the probabilities (11) over the set $\mathcal{C}[ 0, 0, [AB], t]$ of all the trajectories which end up in the interval $[A, B]$, i.e. :

\[
\mathbb{P} \{x(t) \in [AB]\} = \int_{\mathcal{C}[0, 0, [AB], t]}\ \prod_{\tau = 0}^t \frac{dx(\tau)}{\sqrt{4\pi D d\tau}}\ \mathrm{exp} \left\{- \frac{1}{4D} \int_{0}^td\tau \dot{x}^2(\tau)\right\}
\]
\begin{equation}
= \int_{A}^B dx \frac{1}{\sqrt{4\pi Dt}} \ \mathrm{exp} \left\{-\frac{x^2}{4Dt}\right\}
\end{equation}

where the second member follows from (9) and (8).

The symbol :
\[
\int_{\mathcal{C}[0, 0, [AB], t]}
\]

formally denotes the summation over the set of trajectories and, as this set is continuous, one has to use the symbol of an integral. The summation over a set of trajectories of the type (12) is called the {\it Wiener path integral}.

So, starting from the Brownian transition probability and distribution, one has quite naturally arrived at the measure and integral over the {\it functional infinite-dimensional space} of all trajectories $x(\tau)$.

Concerning the properties of this set of functions, Wiener has proved that, in the case of path integrals with measure, the set $\mathcal{C}$ of functions which contribute to the integral consists of continuous but non-differentiable functions. Functions are continuous because the paths are continuous at $t = 0$ and hence, at any moment $\tau\ (0 \leq \tau\leq t$), due to the homogeneity of the Brownian process in time. However, considerations of the continuous limit for the Brownian motion has shown that the velocity for the Brownian particle is ill-defined. So the path integrals, contrary to our intuition and to what our graphs seem to suggest, are not smooth but sums over fully "zigzag-like" trajectories, corresponding to non-differentiable functions, described more precisely now in the framework of fractal theory.

Let us achieve our comment on Wiener path integral with an important point, very crucial for the following : Wiener has proved the following theorem :\\

\begin{thm}[Wiener]
In the case of Brownian motion, the Wiener path integral is equal to zero over both the set of discontinuous and the set of differentiable trajectories. In more precise mathematical terms, the set of continuous as well as the set of differentiable functions have a zero Wiener measure.
\end{thm}

\subsection{Feynman and Wiener path integrals}

Returning now to the theory of Feynman path integral, we can see that there is a formal link between the two types of path integrals - Feynman's and Wiener's - because, while the Schrödinger equation of a massive particle non-relativistic free is written:
\[
 -{\frac {\hbar ^{2}}{2m}}\ \Delta \psi ({\vec {r}},t)\ =\ i\hbar \ {\frac {\partial \psi ({\vec {r}},t)}{\partial t}},
  \]
 where $\psi$ is the quantum wave function, the diffusion equation in the space for the density probability $P$ is :
  \[
  D\ \Delta P({\vec {r}},\tau )\ =\ {\frac {\partial P({\vec {r}},\tau )}{\partial \tau }}.
     \]
   
   So it is enough to put $D = - \hbar / 2m$ for the diffusion coefficient, and $t = i \tau$ for the time, to transform the Schrödinger equation into a diffusion equation. Now, it turns out that the Wiener path integral - for the diffusion equation - is easier to define mathematically rigorously than that of Feynman - for the Schrödinger equation. Some authors have therefore proposed to define the Feynman integral from Wiener's measure by making an analytic extension for imaginary times.

Unfortunately, it is not so easy to do so, and moreover, this is not sufficient to solve the problems of the Feynman path integral, particularly the fact that, unlike Wiener's, it is not a «measure» in the full mathematical sense of the term. It seems that theses difficulties have already been seen in the history of science, to the point that some have forbidden to take the step daringly made by Feynman..

\section{Bachelard and Buhl on micro-trajectories}
	
	At the 4th chapter of his book {\it The Philosophy of the No}, the French epistemologist Gaston Bachelard comments on a memoir by the mathematician Adolphe Buhl, entitled "On some corpuscular and wave analogies", published in 1934 in the French {\it Bulletin of mathematical sciences} (\cite{Buh}). 
	
	Adolphe Buhl (1878-1949) was a self-taught mathematician. At the age of 14, a paralysis immobilizing him for a few years and forcing him to walk on crutches for the rest of his life, he became interested in mathematics, reached a high level of expertise, and obtained a PhD in 1901 at the Faculty of Sciences of the University of Paris. His main thesis was about differential equations and partial derivative ones. His complementary thesis was about astronomy (the Delaunay theory concerning the movements of the moon).The thesis committee was composed of Gaston Darboux, Henri Poincaré and Paul Appell, some of the best mathematicians of the time.
	
	Around the 1930s, mathematicians knew that the corpuscular space, as Louis de Broglie had shown, could not be that of the "figures and movements" of classical mechanics. Several of them had tried to find forms of geometry adapted to the microphysical world : channel space (N. Bohr), infinitesimal geometry and geometric uncertainty schemes (G. Bouligand), isometric trajectories on the circle (G. Humbert and M. d'Ocagne). Reasoning on these last representations, which were the simplest, A. Buhl introduced first the following analytical considerations:

	Let $C$ be a circle, of center  $O$ and radius $a$, on which we consider two arbitrary rays $OA$ and $OB$, separated by an angle $\theta$. The curves $s$ on which the $OA$ and $OB$ rays intercept an arc equivalent to the circular arc $AB$ are given by the differential equation:
\[
 ds = ad\theta \quad \textnormal{or} \quad dr^2 +  r^2d\theta^2 = a^2d\theta^2,
\]
whose general integral is :
\[
 r = a\cos (\theta - c).
\]
\begin{figure}[h] 
	   \centering
	      \vspace{-4\baselineskip}
	   \includegraphics[width=4in]{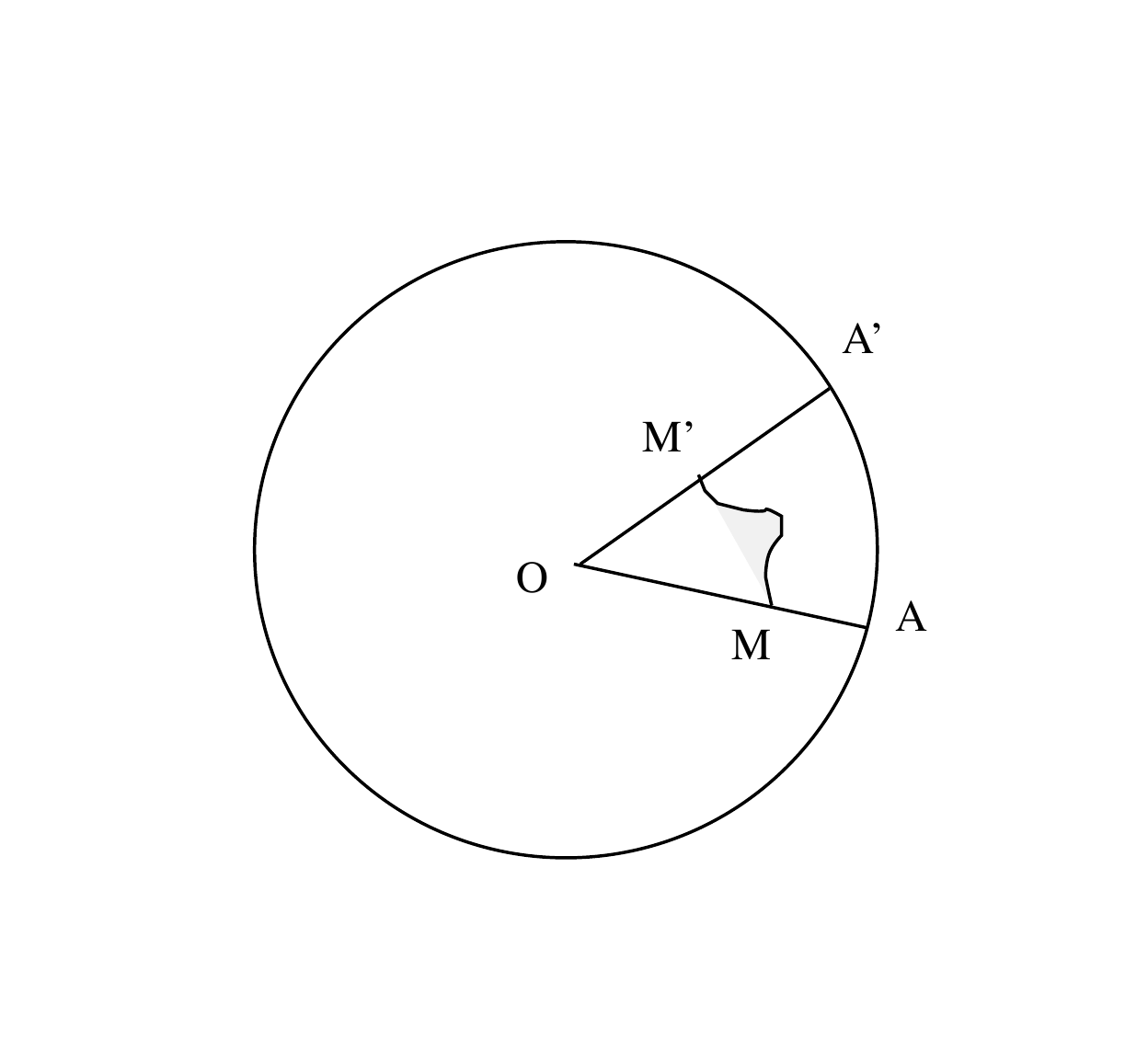} 
	      \vspace{-2\baselineskip}
	   \caption{A problem of integration}
	   \label{fig: Bach1}
	\end{figure}

It represents all the circles, of diameter $a$, passing by $O$ thus tangents internally to the given circle. Through any point $M$, inside the latter, we can pass two circles (2) symmetrical to each other with respect to the common cord $OM$ The given circle, an envelope of circles (2), is singular solution of the differential equation.	 

\begin{figure}[h] 
	   \centering
	      \vspace{-2\baselineskip}
	   \includegraphics[width=4in]{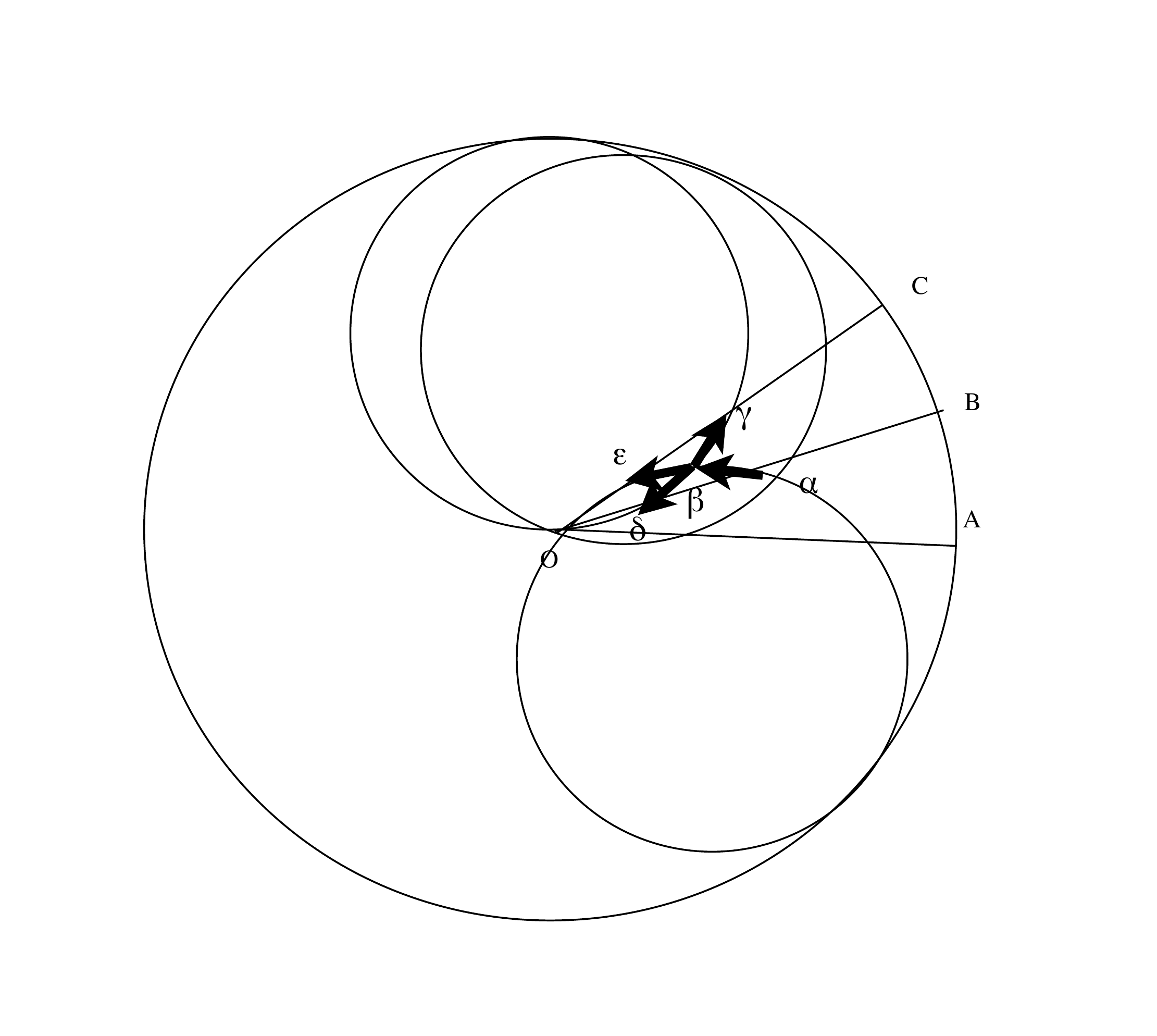} 
	      \vspace{-2\baselineskip}
	   \caption{Buhlian trajectories}
	   \label{fig: cercles1}
	\end{figure}

Buhl comments on his graph as follows :

«Thus, let $OA,\ OB,\ OC$ be neighboring rays... I leave from $\alpha$ on $OA$ with either arcs $\alpha\beta$ or $\alpha\gamma$ equal to AB. 
Here, I am not forced to analytically continue the arc $\alpha\beta$ in $\beta\delta$.
I can go on  continuously with $\beta\epsilon$ and, in $\epsilon$, indeed, an analogous choice will still be possible. And so on. Such a path
than $\alpha\beta\epsilon$... when the elements $\alpha\beta,\ \beta\epsilon$... will become infinitely small, when at the same time $AB,\ BC, ...$ does not cease, for an indefinitely subtle observer who always has the generation in mind, to satisfy the initial statement. (...) 

«There is an indeterminate infinity of vulgar curves which could be drawn on the figure with a pencil always giving a line of a certain width, a line in the thickness of which a subtle observer can inscribe a path $\alpha\beta\epsilon$... That's how a rough curve may have a fine structure or, more accurately, an indefinitely fine structure that solves some physical questions (...)

«A path with angular points, such indefinitely close together, also takes the appearance of the broken line, with infinitely small elements, which is one of the possible representations for the continuous function without derivative (...) The Problem of Determinism can still find here stresses developments. The path $\alpha\beta\epsilon$ is in no way determined by his first element $\alpha\beta$. Where are we going away, starting from $\alpha$? This is not decided yet. This is exactly the only answer we have when one wonders about the behavior of an electronic trajectory.

The uncertainties of Heisenberg can also be easily illustrated in the order of ideas discussed here. On a path in teeth of sawing, with an indefinitely fine structure, a more precise location more and more prevents the design of a tangent. And, on the whole path, the tangents parallel to a given direction tend to become infinitely numerous, so that a tangent direction does not locate a position.» (\cite{Buh}, 338-339).

Choosing circular sections rather than straight lines for the particles path had been considered by Feynman and Higgs in their 1965 book, where they wrote:

«It is possible to define the path in a somewhat more elegant manner. Instead of straight lines between the point $i$ and $i+1$, we could use section of the classical orbit. Then we could say that $S$ is the minimum value of the integral of the lagrangian over all the paths which go through the specified points $(x_{i}, t_{i}$. With this definition, no recourse is made to arbitrary straight lines.» (\cite{Fey2}, 34).

By generalizing his perspective to space - the "space of paths" becoming, in his language, the "space of the canals" - Buhl finds for $S$ (what will be called, after Feynman, the "propagator" ) an expression of Green's function.

From the above quotations, we may see that Buhl was in fact very close to discovering (or rediscovering, since it was already known by Wiener and implicit in his work on Brownian motion) the notion of «path integral». If Buhl does not do so, it is, on the one hand, because he observed, like Wiener, that the measure of the functional expressions is null, and, on the other hand, because he knew that such an integration would exceed the analytical possibilities. In particular, he writes: 

	«What is achieved with these theoretical considerations ... these are the elements in propagation (masses, loads, etc.) (...) To go further, to perceive a true geometry of the propagation, it would be precisely necessary to integrate ... But this is beyond analytical and deterministic possibilities.» (\cite{Buh} 343), 

	The rest of Buhl's text is less prophetic because it seems more or less to give up developing a representation of particle motions and trajectories, especially since, as he still observes, the total function may depend on an infinity of settings.  He noted, however, that Schrödinger's equation  «corresponds to what is most accessible in the theory of wave or corpuscular propagation in channel spaces.» (\cite{Buh}, 345).
	
	The introduction of time, as well, seemed to have caused him many problems, to the point of forcing him to invent a hypothetical «hypertime» to try to stay as close as possible to the known geometry. Ultimately, he had to admit defeat and to recognize that some representations of new microphysics are in fact, as he says, «ageométric». Feynman, as we know, will ignore, even if the mathematical justification of his transgressive gesture has to wait a few years.
	
Bachelard, for whom the memoir of Buhl is an example of rectification of what he calls «the first intuitions», did not fail to praise Buhl. «The ingenuity of Buhl's memory is really integrating ambiguity all along the integral curve» (\cite {Bac}, 97-98). Better: «with respect to the lines $OA$ and $OB$ taken as traces of a wavefront, the family of the Buhlian trajectories constitutes the {\it set of possible paths} for light rays» (\cite{Bac}, 102). 

Finally, Bachelard not only notes the possibility of integrating, alongside families of normal trajectories, indefinitely fine ones (what one would call today fractal trajectories) or others possibly folded on themselves. He concludes by showing that Buhl's text operates a sort of rationalization of the Heisenberg principle.

Of course, there is still no question, in these texts, of explicitly defining an "integral on all paths", and neither Buhl nor Bachelard think of Dirac's article that allowed Feynman to formally deduce his algorithm from the Lagrangian. But this kind of views, in which the usual considerations of Schrödinger's function or the Heisenberg's relations of uncertainty give way to problems of path, have for us, after Feynman's work, a familiar air. Buhl and Bachelard have had a certain intuition of the importance of this representation, without, however, going to the formalism that would have allowed it to be effective. Buhl, in particular, judged it mathematically impossible.

\section{Problems and solutions}

	The problems on which Buhl had stumbled and which had dissuaded him from going further would take time to be solved.
	
	After the intuitive breakthrough of 1948, pedagogically repeated in the Feynman and Higgs 1965 book, the subsequent history of the path integral (see \cite{Klau} shows that many authors – see the more formal approaches of Feynman and Kac (1951), Gel'fand and Yaglom (1956), Cameron (1960), Itô (1962), Daubechies and Klauder (1985) and so on – tried desperately to make rigorous what was not.

	While recognizing the heuristic nature of Feynman's method, many mathematicians have, moreover, confessed their embarrassment. Typical of the genre are, for example, remarks by Yen Chin Ong of National Taiwan University, who presents (with humour) his statement as "mathematician's lament":

	«Despite the successfully predicting power of Feynman path integral, it lacks mathematical rigor. Trained as a mathematician, I have difficulty accepting the validity of path
integral, and for that matter, most of quantum field theory; although as a physicist, I know how to use them and to wave my hands as necessary, deep down I am deeply troubled»(\cite{Ong}, 1-2).

	In his note, Ong did not fail to point out everything that seemed shocking to a mathematician. Rewriting the path integral as :
	\[
\int_{\Gamma} D\gamma
\]
	where $\gamma : [t_{i}, t_{f} \to \mathbb{R}^d$ is any path connecting the endpoints $\gamma(t_{i}) = q_{i}$ and  $\gamma(t_{i}) = q_{i},\ \Gamma$ being the space of paths, he recalled that $D_{\gamma}$ should have been thought as a Lebesgue measure on $\Gamma$, which could not be done, since this Lebesgue-type measure {\it simply does not exist}.
	
	As Ong explained, «this follows from the well-known result in functional analysis that a [nontrivial] translational invariance Lebesgue- type measure cannot be defined on infinite dimensional Hilbert spaces» (\cite{Ong}, 2). In the case of Brownian motion (the so-called Wiener process in the one-dimensional version), the Wiener measure is not translationally invariance and, in 1960, Cameron proved that it was not possible to construct «Feynman measure» as a «Wiener measure» with a complex variance, i.e. as limit of finite dimensional approximations of the expression:
\[
\frac{e^{(i/\hslash)} \int_{0}^t (m/2) \dot{\gamma}(s)^2ds D\gamma}{\int e^{(i/\hslash)} \int_{0}^t (m/2) \dot{\gamma}(s)^2ds D\gamma}.
\]
On the contrary, the usual Lebesgue measure on $\mathbb{R}^d$  has finite total variation on bounded measurable subsets of $\mathbb{R}^d$. All this, of course, is very well-known and many discussions on the attempts to make mathematical sense of the path integral formulation have result (see \cite{Gro1}, \cite{Gro2}). Feynman himself knew this lack of rigor and hope that, as for the attemps of Cavalieri in the period that preceded the development of the differential calculus, one would be finally led to some more rigorous formulation. 

It seems that this time has now come. The rigorous, long-awaited formulation has finally arrived.

After having quickly located the origin of the path integral in the works of Wiener, Dirac and Feynman, Pierre Cartier and Cécile de Witt-Morette, in the major book which they devoted to the subject, underlined the difficulty, noticed by Mark Kac, related to the Feynman integral, the presence of the complex number $i$ in the exponent of the formula of integration not making easy its rigorous expression (see \cite{Car}, 4).

They then ask the three leading questions related to the introduction of this formalism so necessary and yet so problematic that it made say to Simon Simon that it was an «extremely powerfull tool used as a kind of secret weapon bby a small group of mathematical physicists» (\cite{Sim}) :

\begin{enumerate}
\item How does one choose the short-time probability-amplitude $(q'_{t+\delta t}| q'_{t})$ and the undefined normalization constant?
\item How does one computes the $N$-tuple integral?
\item How does one know whether the $N$-tuple integral has a unique limit for $N = \infty$?
\end{enumerate}

The answer, they wrote, «is to do away with the $N$-tuple integrals and to identifiy the function spaces which serve as domain of integration for functional integrals. The theory of promeasures (projective systems of measures on topological vector spaces, which are locally convex but not necessarily locally compact) combined with Schwartz distributions yields a practical method for integretating on function spaces» (\cite{Car}, 6). Once having crossed the step from promeasures to prodistributions, these ones are used to solve the main problems and a useful byproduct or prodistributions is presented with the definition of complex  gaussians in Banach spaces.

As we can see, a non-trivial mathematics has proved necessary to make Feynman's intuitions rigorous.

Following a presentation of Pierre Cartier (see{Car2}, 53-54), rather than entering into this very heavy technique, we will confine ourselves here to bringing back the spirit of this discovery, insisting on the change of point of view that assured the authors the success of their approach.

The three questions posed above, in fact, are summarized in one, which testifies to the major problem posed by the Feynman integral: when we move from classical mechanics to quantum mechanics, $\hslash$ is the only new element which intervenes and the integration element $D$ of the functional $X$ representing the paths comprises an infinity of variables. Therefore, when we approximate this functional space to an infinity of dimensions by a space with a finite number of dimensions, replacing $X$ by a finite data $(x_ {1}, ..., x_ {N-1}$ a normalization constant is used. We then replace the $DX$ by $d_ {x_ {1}}, ..., d_ {x_ {N}}$, we integrate and we go to the limit. Feynman pretended to believe that this procedure was unambiguous. Of course it is not the case.

It is not the case because it is only in finite dimension, in a space such that $\mathbb{R}^N$, that an element of volume $d^N x = d_ {x_ {1}} , ..., d_ {x_ {N}}$ can verify this property of the notion of measure which states that, for any homotethic ratio $c$, the unit of volume will multiply by $c^N$, that is, we will have:
\[
d N (cx) = c Nd Nx.
\]
Indeed, in infinite dimension, that is to say for $N$ infinite, we have only three possibilities:
\[
c ^ N = 0 \qquad \textnormal {if} \ c <1;
\]
\[
c ^ N = 1 \qquad \textnormal {if} \ c = 1;
\]
\[
c ^ N = + \infty \quad \textnormal {if} \ c> 1;
\]
In other words, the fact that the coefficient of proportionality $c^{\infty}$ is infinite (or zero) is equivalent to saying that the density of a measure relative to another is infinite, which simply means that, when we make a change of scale, that is to say, a homothety, we change totally sector and we find ourselves in a theory that is his relation to the previous one. In other words, $DX$ is not unique at all.

Contrary to most attempts to solve the problem of the Feynman integral, the solution proposed by Cartier and DeWitt-Morette does not aim to find from the outset a method of crossing to the limit from the finite dimension. It consists first and foremost in establishing an axiomatic framework which ensures the uniqueness of the measure sought, the approximation processes and the actual calculations - whatever they may be - then being a priori legitimized.

The process starts from the examination of the conditions to be met in order to reach the desired goal.

We assume at first a Hilbert space $\mathcal {H}$ and we assume that we want to establish in $\mathcal{H}$ an integration element $Dx$, invariant by translation. The first thing to do, from this perspective, is to try to normalize:
\[
\int_{\mathcal{H}} e^{-\pi x^2} Dx = 1.
\]
D'après l'invariance par translation $D(x_{0} +x) = Dx$, on aura :
\[
\int_{\mathcal{H}} e^{-\pi x^2 -2\pi x_{0}} Dx = e^{-\pi x_{0}^2}.
\]
A reasonable integration element $Dx$ must satisfy this equation. For technical reasons, it is advantageous to take a complex number $x_ {0}$, that is to say an element of the complexified of the real Hilbert space $\mathcal{H}$, which makes it possible to use the Fourier transform. The general formulation of the condition above gives:
\begin {equation}
\int_{\mathcal {H}} \Theta (x | J) Dx = Z (J),
\end {equation}
where $J$ is the complex number $x_ {0}$ and $ Z (J) = e^ {\pi x^ 2_{0}}$.

The benefit of the operation is that if we set $\Theta (x |.)$ And $Z (.)$, there is at most one integral that checks (13), and this, {\it whatever J is}.

The question then is: what functions can we integrate? And the answer of Cartier and DeWitt-Morette, following Albeverio and Hoegh-Krohn (see \cite{Alb}), was as follows:

Suppose we have a $F(x $ function that reads as follows:

\begin{equation}
F(x) = \int_{\mathcal{K}} \Theta(x|J) d\mu(J),
\end{equation}
where $x$ is in a certain Hilbert space $\mathcal {H}$ and $J$ in another Hilbert space noted $\mathcal {K}$ (a parameter space). Suppose then that $\mu$ is a complex measure of finite mass, well defined. The equality (14) corresponds to a generalization of the Fourier transform and the functions $F(x)$ are generalizations of the complex Fourier-Stieltjes transforms (in other words, a Fourier transform of the measures). The axiomatic program of Cartier-DeWitt-Morette is to generalize this situation.

They are given two Banach spaces $\mathcal{H}$ and $\mathcal {K} $ (usually Hilbert spaces, a $\Theta (x)$ function with continuous complex values and another function $Z ( J)$ only complex, and it is a question of finding an integrator $Dx$ for which the preceding properties are checked. It will be, in fact, a class of test-functions and an integral for them. The test-functions will be the functions $F (J)$ of the form (14) with a complex Borelian measure $\mu$ , bounded on $\mathcal {K}$. The norm of $F$ will then be the total variation of $\ mu$, or rather, we will have $|| F || = \mathrm{inf\ var}\ \mu$, where the infimum is taken on all the $\mu$ measures concerned. It is easy to see that all of these functions $F$, equipped with this standard, is a Banach space, so the natural definition of the integral of $F$ is:

\begin{equation}
\int_{\mathcal{H}} F(x) Dx = \int_{\mathcal{K}} Z(J) d\mu(J).
\end{equation}

We then have almost automatically a theorem of uniqueness and the existence theorems are also easily demonstrated: they consist in showing that we can pass to integrals to a finite number of variables and to make some passages to the limit.

Feynman's procedure of the path integral is then justified, at least in the sense that for each value of $c$ we have a uniqueness, which is enough to give the solution of a number of paradoxes of this theory of integration in infinite dimension.

Is it sufficient to solve all the problems? Must we think the game is over? It is not sure.

It can be argued (see \cite{Mon}, 7)  that the Cartier-DeWitt-Morette solution is not completely satisfactory: 

\begin{enumerate}
\item The quantum dynamics is described by a generalized measure which is the product of the complex exponential of the action and the Lebesgue-Feynman measure times the determinant, so that the non-invariance of the dynamics with respect to a transformation cannot disappear.
\item There still remains also the question of clarifying the role of renormalization in the problem of quantum anomalies.
\end{enumerate}

It remains that a big step has been taken in the way of bringing a mathematical rigor to the integral of Feynman and, even if the file is not completely closed, progress is now sufficient to justify the transgressive gesture of the physicist by ensuring the convergence of its integral in all cases where the functional spaces are well chosen.


\begin{thebibliography}{}\addcontentsline{toc}{chapter}{Bibliographie}
 
\bibitem[Alb 76]{Alb} Albeverio, S and Hoegh-Krohn, R.J., {\it Mathematical theory of Feynman path integrals}, Springer-Verlag, Berlin-Heidelberg-New-York, 1976.

\bibitem[Bac 66]{Bac} Bachelard, G., {\it La philosophie du non, essai d’une philosophie du nouvel esprit scientifique} (1940), Les Presses universitaires de France, 4e édition, Paris, 1966.

\bibitem[Buh 34]{Buh} Buhl, A., «Sur quelques analogies corpusculaires et ondulatoires», {\it Bulletin des Sciences Mathématiques}, première partie, 333-367, 1934.

\bibitem[Car 06]{Car}  Cartier, P. DeWitt-Morette, C., {\it Functional Integration : Action and Symmetries}, Cambridge University Press, Cambridge, 2006.

\bibitem[Car 12]{Car2} Cartier, P., «L'intégrale de chemins de Feynman : d'une vue intuitive à un cadre rigoureux», in J.-P. Kahane, P. Cartier, et al., Leçons de mathématiques d'aujourd'hui, volume 1, 3e ed., 27-58, Cassini, Paris, 2012.

\bibitem[Cha 01]{Cha} Chaichian, M., Demichev, A., {\it Path Integral in Physics}, Vol. 1, Stochastic processes and quantum mechanics, IoP (Institute of physics publishing), Bristol and Philadelphia, 2001.

\bibitem[Dir 33]{Dir}  P. A. M. Dirac, «The Lagrangian in quantum mechanics», {\it Phys. Zeit. der Sowjetunion} 3,  64-72, 1933.

\bibitem[Ein 05]{Ein1} Einstein, A., «Über die von der molekularkinetischen Theorie der Wärme geforderte Bewegung von in ruhenden Flüssigkeiten suspendierten Teilchen«, {\it Ann. der Phys.}, 17, 549-560, 1905.

\bibitem[Ein 06]{Ein2} Einstein, A., «Zur Theorie der Brownschen Bewegung», {\it Ann der phys}, 4, 19, 371-381, 1906.

\bibitem[Fey 48]{Fey1} Feynman, R. P., «Space-time approach to non-relativistic quantum mechanics», {\it Review of Modern Physics} 20, 2, 367-387, 1948. 

\bibitem[Fey 65]{Fey2} Feynman R. P. and Hibbs, A. R., {\it Quantum Physics and Path Integrals}, McGraw-Hill, New York, 1965.

\bibitem[Gro 93]{Gro1} Grosche, C., «An introduction into the Feynman path integral», {\it ArXiv:hep-th/9302097v1}, 20 Feb 1993.

\bibitem[Gro 98]{Gro2} Grosche C., Steiner F., {\it Handbook of Feynman path integrals}, Springer Verlag, Berlin, Heidelberg, 1998.

\bibitem[Klau 03]{Klau} Klauder, J. R., «The Feynman Path Integral: An Historical Slice», {\it ArXiv:quant-ph/0303034v1}, 1-24, 7 Mar 2003.

\bibitem[Mon 16]{Mon} Montaldi, J., Smolyanov, O. G., «Feynman path integrals and Lebesgue-Feynman measures», {\it ArXiv: 1612.06657v1 [maht-ph]} 20 Dec 2016.

\bibitem[Ong 18]{Ong} Ong, Y. C., «Note: Where is the Commutation Relation Hiding in the Path Integral Formulation?», $https://physicstravelguide.com/\_$media/quantum... /path-integral.pdf, 1-6, 15 mai 2018.

\bibitem[Sau 08]{Sau} Sauer, T., «Remarks on the origin of path integration, Einstein and Feynman», {\it ArXiv:0801.1654v1  [physics.hist-ph]}, 1-11, 10 Jan 2008.

 \bibitem[Sch 86]{Sch} Schweber, S. S.,  « Feynman’s visualization of space-time processes », {\it Review of Modern Physics}, vol. 58, no 2,? 449-508, 1$^rst$ avril 1986.
 
 \bibitem[Sim 79]{Sim} Simon, B., {\it Functional Integration and Quantum Physics}, Academic Press, Cambridge (Mass.), 1979.
 
 \bibitem[Wie 21$^1$]{Wie1} Wiener, N., «The Average of an Analytic Functional» PNAS 7 (9), 253-260, September 1, 1921.

\bibitem[Wie 21$^2$]{Wie2} Wiener, N., «The Average of an Analytic Functional and the Brownian Movement», PNAS 7 (10), 294-298, October 1, 1921.
 
 \bibitem[Wie 23]{Wie3} Wiener, N., «Differential space», {\it J. Math. Phys.}, 2, 131-174, 1923.
 
\bibitem[Wie 24]{Wie4} Wiener, N., «The Average value of a Functional», Proc. of the London Math. Soc.,  Volumes 2-22, Issue1, 454-467, 1924.
 
\end{thebibliography}
\end{document}